# Harnessing coupled nanolasers near exceptional points for directional emission

Guilhem Madiot[1,*], Quentin Chateiller[2], Alexandre Bazin[2], Patricia Loren[1], Konstantinos Pantzas[2], Grégoire Beaudoin[2], Isabelle Sagnes[2], and Fabrice Raineri[1,2]

[1] Université Côte d'Azur, CNRS, Institut de Physique de Nice, 06200, Nice, France
[2] Centre de Nanosciences et de Nanotechnologies, CNRS, Université Paris-Saclay, 10 Bd Thomas Gobert, 91120 Palaiseau, France
* gmadiot@unice.fr

## Abstract

Tailoring the losses of optical systems within the frame of non-Hermitian physics has appeared very fruitful in the last few years. In particular, the description of exceptional points (EPs) with coupled resonators have become widespread. The on-chip realization of these functionalities is highly significant for integrated nanophotonics, but requires fine control techniques of the nanodevice properties. Here, we demonstrate pump-controlled directional emission of two coupled nanolasers that distantly interact via an integrated waveguide. This coupling scheme unusually enables both frequency- and loss-couplings between two cavities, which can be advantageously exploited to reach EPs by either detuning the cavities or controlling the gain of nanolasers. The system can be readily reconfigured from bidirectional to unidirectional emission by adjusting the pump power.

## Introduction

Exceptional points (EPs) are degeneracies in the complex eigenspectrum of coupled resonators where at least two eigenvalues coalesce, with their respective eigenvectors [1–3]. Their presence is associated to a rich phenomenology including spatial [4–6] or spectral [7,8] nonreciprocity, topological singularities [9–12], or enhanced sensitivity [13–17]. In practice, to reach an EP, the coupling rate must be exactly compensated by an asymmetry in the resonators' natural properties, i.e. frequencies and loss rates. Hence, this operation requires a fine control over the former and/or over the coupling constant itself. Most demonstrations exploit an evanescent coupling between two resonators, leading to a pure energy repulsion, i.e., a real splitting between the resulting Quasi Normal-Modes (QNMs). In this case, an EP can only be reached by introducing an imbalance between the resonators' dissipation rates. Meanwhile, their frequencies must be perfectly tuned. This situation defines the well-known "gain-loss" configuration, where gain is anti-symmetrically distributed within two resonators with identical natural frequencies, enabling Parity-Time (PT) symmetry to be broken or restored by crossing an EP [18–20].

Exceptional points can also be reached if the coupling has a dissipative contribution, i.e., if the coupling constant is complex in the non-Hermitian physics formalism. Few examples include e.g., multimode cavity optomechanics, where the light-mediated effective coupling between two micromechanical resonators leads to both frequency- and loss-splitting in the eigenspectrum [8,21–23]. Several works have shown that a loss-splitting can be obtained in microring resonators, by coupling the clockwise and counter clockwise modes via a scattering mechanism [4,5,24].

Loss-splitting has also been recognized for a long time as an intriguing consequence of waveguide-coupled standing-wave cavities as described by coupled mode theory [25–27]. The coupling rate between two cavities interacting via a waveguide, as depicted in Fig.1A, takes the form $K = \Gamma_c e^{j\phi}$. Here, $\Gamma_c$ is the cavity-to-waveguide coupling rate, and $\phi$ is the phase acquired by the light between the cavities, referred to as the coupling phase in the following. Assuming an independent cavity decay rates, $\Gamma_A$ and $\Gamma_B$, and natural angular frequencies, $\omega_A$ and $\omega_B$, the coupled-mode analysis of this configuration provides the rate equations for the cavity amplitudes, $\mathbf{A} \equiv (a_A \quad a_B)^T$:

$$j\frac{d\mathbf{A}}{dt} = \begin{pmatrix} \omega_A - j\Gamma_A & -jK \\ -jK & \omega_B - j\Gamma_B \end{pmatrix} \mathbf{A} \quad (1)$$

In the general case, the properties of the QNMs in this system are given by the complex eigenvalues of the above effective Hamiltonian. They involve the cavities' average properties $\overline{p} = \frac{1}{2}(p_A + p_B)$, and dissimilarities, $\delta p = \frac{1}{2}(p_A - p_B)$:

$$\Lambda_\pm = \overline{\omega} - j\overline{\Gamma} \pm \sqrt{(\delta\omega - j\delta\Gamma)^2 - K^2} \quad (2)$$

By setting the complex splitting to zero in Eq.2, two EPs are found at positions $\{\delta\omega_{EP}=\pm\Gamma_c\cos\phi, \delta\Gamma_{EP}=\mp\Gamma_c\sin\phi\}$ of the parameter space, as illustrated in Fig.1 B. It must be noted that the effective Hamiltonian in Eq.1 is generally not PT-symmetric (9,10), unless the cavities are set in the "gain-loss" configuration mentioned above, i.e. displaying EPs when $\omega_A=\omega_B$, $\Gamma_A = -\Gamma_B = \Gamma_c$, and with $\phi = \frac{\pi}{2}[\pi]$. Interestingly, in the proposed modal configuration, the system can be drastically reconfigured by a simple modification of the inter-cavity distance. For example, setting the phase shift to a multiple of $\pi$ enables the EPs to be reached by varying the frequency detuning rather than the gain.

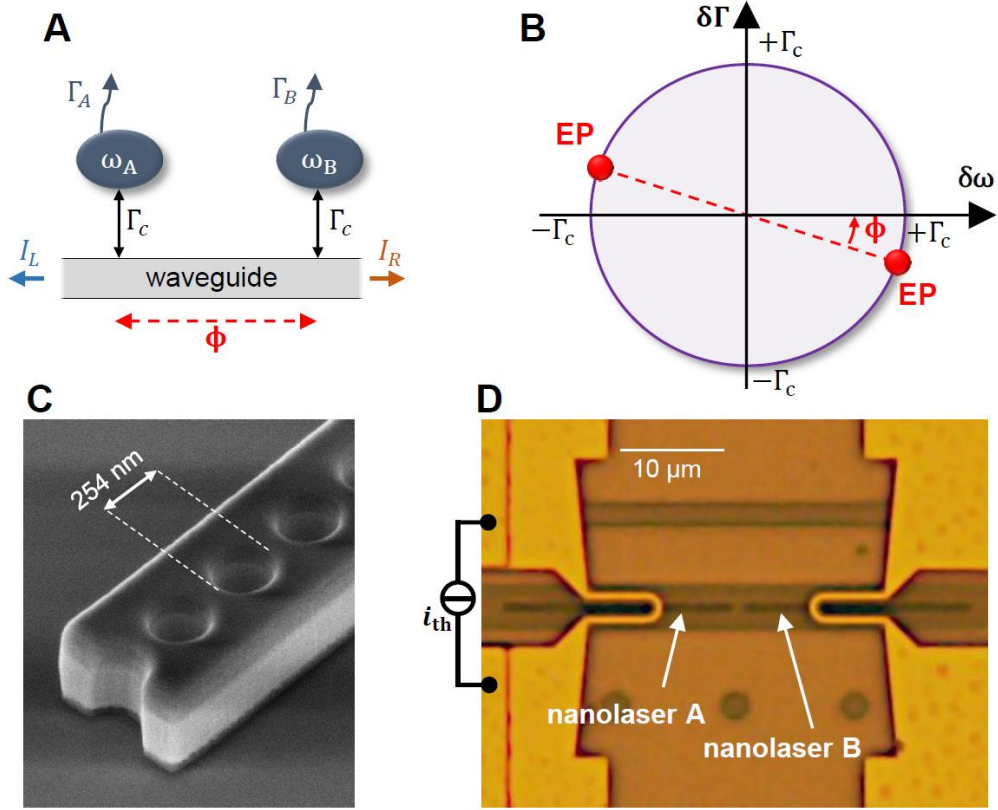

**Fig. 1. Exceptional points (EPs) with waveguide-coupled cavities.** (**A**) Two waveguide-coupled nanocavities with independent loss rates and frequencies. (**B**) Localization of the EPs in the parameter space $\{\delta\omega, \delta\Gamma\}$, for a given coupling phase $\phi$ (**C**) SEM micrograph of an InP photonic crystal nanolaser. (**D**) Microscope image of the coupled nanolasers with integrated thermoresistive gold nanowires used to tune the nanolasers frequencies.

## Results

In order to implement this scheme, we employ a pair of waveguide-coupled nanolasers. Each laser is formed by an indium-phosphide photonic-crystal nanobeam cavity [28] and is integrated over a silicon waveguide (9). A scanning electron microscopy (SEM) image of a single nanolaser is shown in Fig.1 C. The cavity-to-waveguide coupling strength, $\Gamma_c$, is determined by both the waveguide geometry, and the cavity-to-waveguide gap [29]. Meanwhile, the coupling phase $\phi$ can be tuned by finely controlling the separation distance between the nanolasers. Based on numerical evaluation of the waveguide effective index, $n_{\text{eff}} \approx 2.3$, we fabricate a chip where the coupling distance is incremented by 35 nm. This is equivalent to an increment of the coupling phase by approximately $\pi/10$. Therefore, we are capable of exploring different coupling regime from purely dispersive to purely dissipative coupling.

Each nanolaser is optically pumped at room temperature by focusing a continuous wave 1064 nm laser diode at its center. Both ends of the underneath waveguide are terminated by a grating coupler, and are addressed with two optical fibers to collect the emission and perform a spectral analysis. In order to precisely control the cavity detuning $\delta\omega$, a gold resistive nanowire is fabricated 1 µm above each nanolaser with SiO$_2$ in between, as shown in Fig.1 D. When current flows, the nanowire heats up, causing a redshift of the cavity mode $\delta_{\text{th}} = \beta_{\text{th}} P_{\text{th}}$ up to 5 nm, where $P_{\text{th}}$ is the heating electrical power.

Applying identical pumping powers on both cavities, we measure the output spectrum as a function of the heating current, which is directly converted into a cavity detuning, $\delta\lambda$, in Fig.2 A-C. The measurement is carried out for three different cavity separation offsets, 315 nm (A), 420 nm (B), and 525 nm (C). In the case A, we observe a splitting of the QNMs, with both dispersive and dissipative components. Indeed, we do not only observe an energy-repulsion but also an asymmetry in their linewidths, such that one QNM becomes narrower and more intense whereas the other one broadens. This is verified by fitting the spectrum with a double-voigt function. In Fig.2 D, we show the real part (green) and the imaginary part (red) of the splitting, i.e. $\lambda_+ - \lambda_-$ and $\Delta\lambda_+ - \Delta\lambda_-$, respectively, with the error bars given by the spectrometer resolution. Fitting the data applying Eq.2 (full lines), we extract, among others, the cavity-to-waveguide coupling rate $\Gamma_c$, and the coupling phase $\phi$. At $\delta\lambda = 0$, we show that the splitting is complex, i.e. both the QNMs spectral positions and linewidths are different. A typical spectrum illustrates this situation in Fig.2 G (top). In the second case (B), we observe a purely dissipative splitting, i.e. one mode broadens while the other gets narrower as approaching $\delta\lambda = 0$. The eigenvalue analysis and fitting shows that an exceptional point (EP) is crossed in this measurement, as pointed out by the red dot in Fig.2 E. Crossing this point is associated with a transition from purely real to purely imaginary splitting. Finally, in the third case, the coupling is fully dispersive, i.e. purely real, translating into the typical avoided crossing of the QNMs in the spectrum whereas their linewidths are relatively identical $\Delta\lambda_+ \approx \Delta\lambda_-$). In order to reach an EP with this structure, one would have to tune the nanolaser frequencies ($\delta\omega = 0$) and ramp the loss-rate difference $\delta\Gamma$. It could be done here by pumping a single nanolaser while keeping the other in an absorption regime [30].

The eigenvalue analysis is repeated for different cavity distance offsets, and we evaluate the mean value of the cavity-to-waveguide coupling rate, $\Gamma_c/(2\pi) \approx 93 \pm 22$ GHz (see Fig.2 H, top). Most importantly, we verify that the phase is proportional to this distance (bottom), where the linear fit provides the effective index of the silicon waveguide, $n_{\text{eff}} \approx 2.1 \pm 0.5$, proving consistent with the expectations as mentioned above.

Directional propagation and emission of light constitute one of the most promising phenomena to be exploited with photonic devices set in the vicinity of EPs [22,31–33]. In our system, the emission directionality results from the dephasing $\phi$ between the waves that couples out from the two cavities, into the waveguide. This dephasing allows interferences that can be destructive on the left side but constructive on the right side, and vice-versa:

$$\begin{pmatrix} s_L \\ s_R \end{pmatrix} = j\sqrt{\Gamma_c} \begin{pmatrix} 1 & e^{j\phi} \\ e^{j\phi} & 1 \end{pmatrix} \begin{pmatrix} a_A \\ a_B \end{pmatrix} \qquad (3)$$

where the left (L) and right (R) output wave amplitudes are defined such that the corresponding intensities write $I_{L,R} = |s_{L,R}|^2$.

In a second experiment, we focus on the directional emission of a pair of interacting nanolasers with the measured coupling parameters $\phi \approx 1.08\pi$ and $\Gamma_c/(2\pi) = 22.7$ GHz. We also determine their respective laser threshold powers independently, $P_{A,t} = 2.2$ mW and $P_{B,t} = 4.3$ mW, as well as the power dependency of the lasers' frequencies and damping rates. This dependency occurs through the depletion of carriers in the semiconductor under pumping. Finally, we calibrate the heating coefficient $\beta_{th}$=1.94 nm/µW, based on which we set the current $i_{th}$ such that the nanolasers have identical natural frequencies. We scan $P_A$ although $P_B$ is ramped progressively, while the output fields are collected on each termination of the waveguide, and detected by two photoreceivers whose output voltages are read with an oscilloscope.

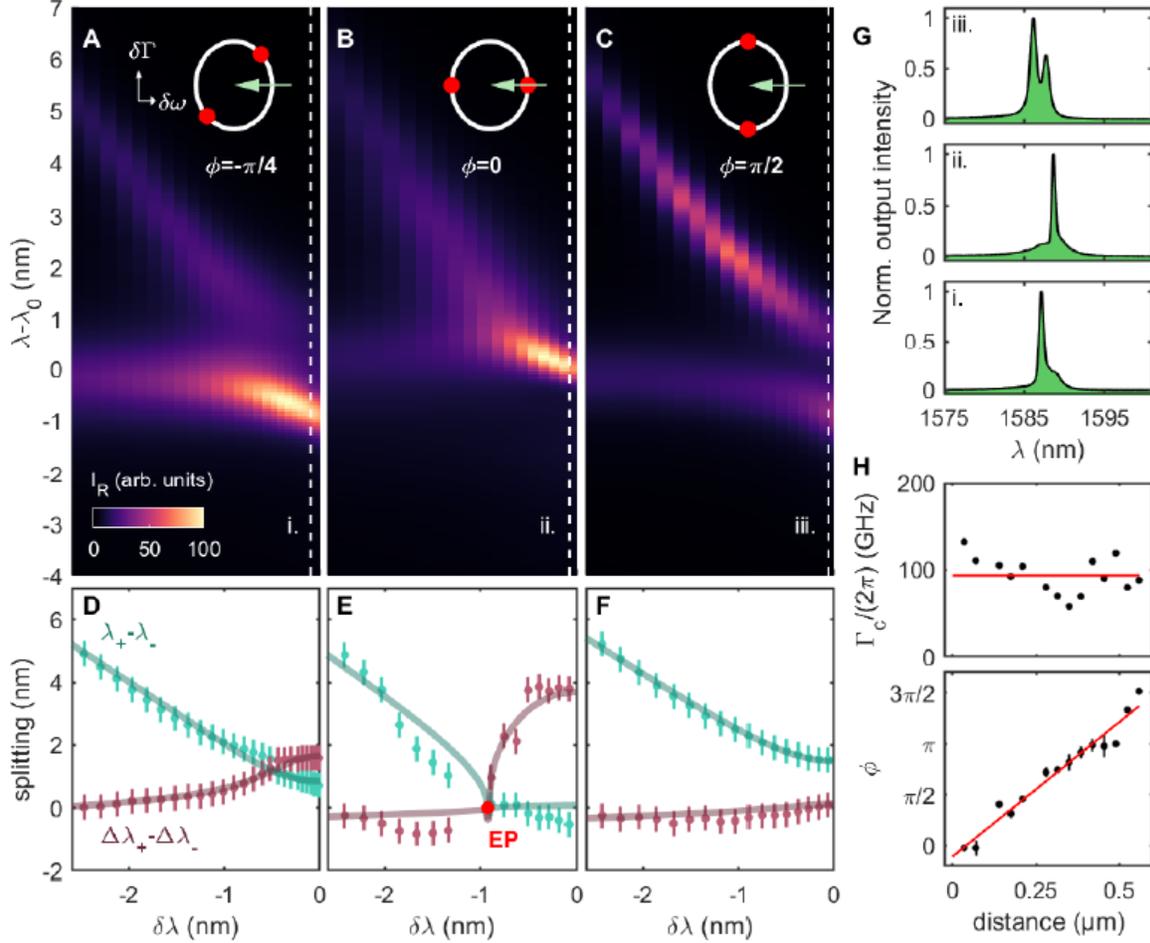

**Fig. 2. Phase-dependent complex coupling.** (**A-C**) Experimental emission spectra as a function of the natural cavity detuning $\delta\lambda$ for different cavity separation lengths. In white, we represent the parameter space $\{\delta\omega, \delta\Gamma\}$ as described in Fig. 1B, and illustrate the trajectory performed experimentally (green arrow) as well as the positions of the EPs (red dots). (**D-F**) Eigenvalues obtained by fitting the upper data. We show the real (green) and imaginary (red) parts of the complex splitting, and fit the data using Eq.2. (**E**) An EP is evidenced (red dot) when $\lambda_+ = \lambda_-$ and $\Delta\lambda_+ = \Delta\lambda_-$ are simultaneously verified. (**G**) From top to bottom: representative spectra are extracted from (A), (B), and (C). (at position indicated by dashed lines) to illustrate purely dispersive, purely dissipative, or complex splitting, respectively. (**H**) The waveguide-coupling rate $\Gamma_c$ (top) as well as the coupling phase $\phi$ (bottom) are both returned by fitting the eigenvalues and shown as a function of the cavity separation length and fitted with a line (red).

In Fig.3 A-B, we compare the left side emission, $I_L$ (A) with the right-side emission, $I_R$ (B). The intensities are shown as a function of the normalized pump powers, below threshold ($P_{A,B}/P_{A,B,t}<1$). On both sides, we note a strong enhancement when the normalized powers are equal due to the system symmetry in this situation, i.e. $\delta\omega = \delta\Gamma = 0$, leading to a strong loss splitting between the QNMs. Therefore, the enhancement of the emission results from the one collective mode whose losses are sufficiently reduced to be overcome by gain, such that it passes the lasing threshold. Though the second QNM's linewidth has broadened and does not significantly contribute to the emission.

When comparing the left and right sides, it appears that the relative intensities vary quite significantly along the map. Therefore, it is useful to introduce the output intensity contrast $C_{out} = (I_L - I_R)/(I_L + I_R)$, as a figure of merit of the emission directionality. It is represented in Fig.3 C where blue and red colors correspond to a dominant emission towards the left side ($C_{out} > 0$) and towards the right side ($C_{out} < 0$) respectively. The balanced bidirectional emission corresponds to white color ($C_{out} = 0$). This map shows an abrupt variation in the output contrast, mostly when the intensity enhances.

In Fig.4 A, we plot the theoretical output contrast in the parameter space $\{\delta\omega, \delta\Gamma\}$, inputting the calibrated experimental parameters in the model. Both EPs (red dots) are represented on the circle $\delta\omega^2 + \delta\Gamma^2 = \Gamma_c^2$. The colormap evidences an abrupt transition from left-side emission (blue lobe) to right-side emission (red lobe) regions. Importantly, the transition between the two EPs corresponds to the path that offers the most contrasted change of directionality. Relying on experimental calibration, we convert the pump powers and the heating current into a position of the system in the parameter space $\{\delta\omega, \delta\Gamma\}$. In particular, in Fig.4 A, we show three different trajectories obtained by sweeping the pump power $P_A$, while $P_B$ still remains constant. These trajectories are dominated by a frequency detuning contribution, which reflects the Henry factor [34] we have calibrated, $\alpha_H \approx 20$. By varying $P_2$ and $i_{th}$, we explore various trajectories and observe different types of responses.

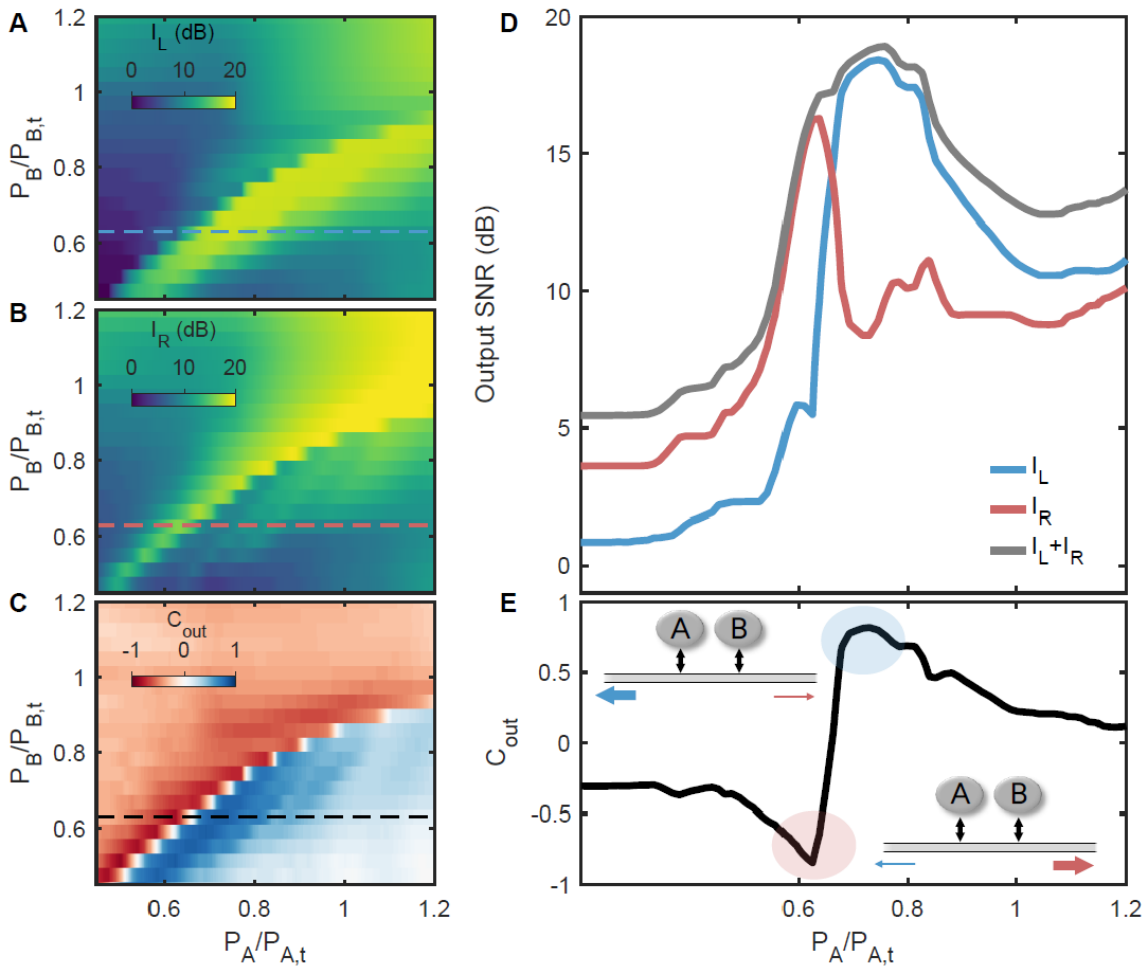

**Fig. 3. Pump-control of the output emission.** (A) Output intensity $I_L$, measured at the left output. (B) $I_R$, measured at the right output. (C) Associated output contrast $C_{out}$. (D) Example measurement of $I_L$ (blue), $I_R$ (orange) and $I_{tot}$, measured at $P_B/P_{B,t} = 0.5$. (E) Associated output contrast showing an abrupt transition from negative values (light emission towards the right output port) to positive values (emission towards the left output port).

The measured output contrast associated to each curve is shown in Fig.4 B, accompanied by the total output intensity in C. The trajectories – labelled i., ii., and iii. – get progressively close to the point {0,0}, where the loss-splitting reaches its maximum. Whereas the response in i. is essentially given by the single nanolaser (A) response, a collective mode builds up in ii. when the nanolaser frequencies cross each other. In iii., the hybridization is maximum, leading to a substantial increase on the QNM response around $P_A/P_{A,t} = 0.4$. Moreover, the output contrast experiences an abrupt inversion, switching from $-0.87$ to $+0.85$ in an interval of only 9% of the threshold power $P_{A,t}$.

The theoretical data shown in Fig.4 A is computed assuming balanced pumping ($\delta P = P_A - P_B = 0$) which is useful to represent the general behaviour of the output contrast in $\{\delta\omega, \delta\Gamma\}$. However, this model does not completely match the experimental conditions where one of the pump powers is swept while the other is fixed, i.e., with a significantly varying $\delta P$. To enable a quantitative modelling of the experiments, the calibrated nanolaser parameters are injected into the coupled mode theory model from which the theoretical output contrast and total output intensities are computed as shown with black curves in Fig.4 B-C. The theory captures well the observations, and the small discrepancies can be imputed to some drifts in the experiment, e.g. of the pump lasers alignments, or of the temperature in the heating nanowires. The coupled mode analysis does not account for the nanolaser nonlinear saturation such that negative decay rate translates into an infinite intensity as observed in Fig.4 C-iii.

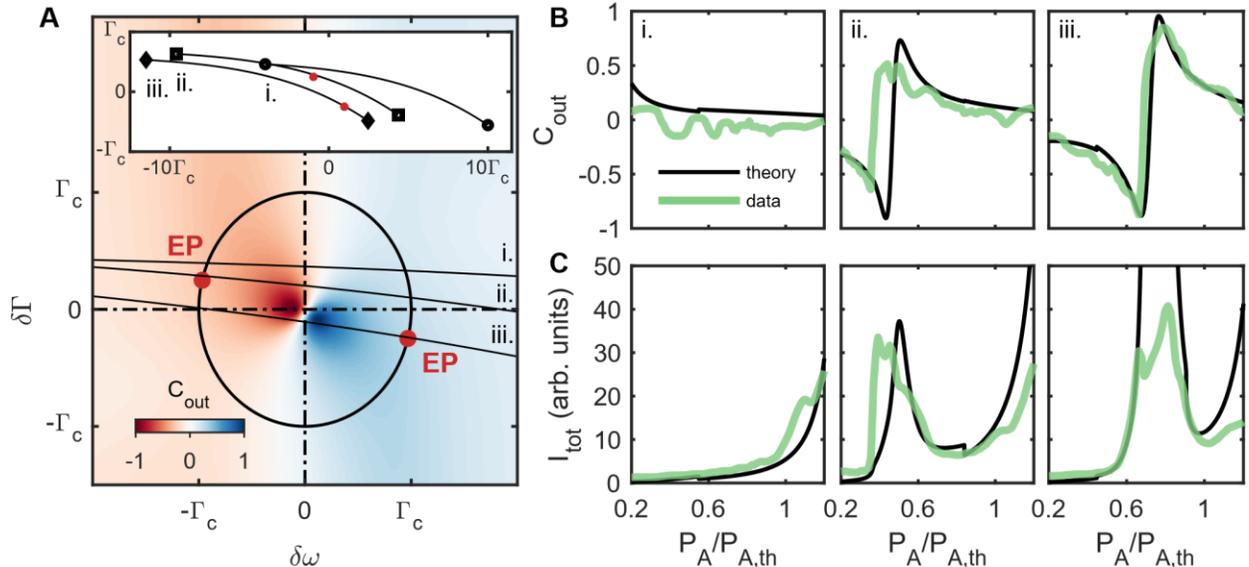

**Fig. 4. Directionality switching near EPs.** (**A**) Theoretical output contrast calculated in the parameter space $\{\delta\omega, \delta\Gamma\}$ with reported positions of the EPs (red dots). Here we assume balanced optical pumping powers. Three trajectories (i., ii., and iii.) are computed based on the experimental calibrations and indicate the path associated to the experimental measurements shown in (B-C). The full trajectories are reported in the inset. (**B**) Output contrast and (**C**) Total output intensity measured as a function of the pump power $P_A$. The measurement (green data) is reproduced at different values of the heating current and of the pump power $P_B$. Theoretical curves are computed based on the nanolaser calibrations (black curves).

## Discussion

We demonstrated that the usual frequency avoided-crossing observed when detuning two coupled resonators turns into a strong loss-splitting if the phase-shift is set to a multiple of $\pi$. The resulting collective lasing emission is characterized by a spatial directionality that can switch depending on the position of the system relative to the EPs, in differential parametric space, $\{\delta\omega, \delta\Gamma\}$. While the emission is perfectly balanced ($C_{out} = 0$) when $\phi = 0[\pi]$ – as directly shown by Eq.3 – the directionality switching is particularly abrupt in the immediate neighborhood of these points. Consequently, these are phase singularities that motivate the implementation of a direct modulation of the phase shift between the nanolasers for their practical utilization [35,36].

The directionality switching bandwidth has yet to be determined but is expected to be limited by the photon lifetime, likely in the range of tens of GHz. In the future, the use of electrical injections [37] would offer an ultimately integrated device in which the directionality could be fully remotely determined. The association of directionality with nonlinear dynamical regimes could be exploited for the realization of all-optical memories [38]. Similarly, self-pulsing or excitable regime encountered in the vicinity of exceptional points would certainly be associated with the spatial nonreciprocity detailed here [39]. More generally, the present platform is an ideal candidate for the exploration of novel laser dynamics occurring in the vicinity of EPs [40–42] and for the join study of nonlinear non-Hermitian photonics.

# Supplemental information

**Sample fabrication process**

We use a SOI chip containing 2.5mm long ridge waveguides with varying widths enabling different coupling strength with the active cavities (*29*). The 280 nm thick InP-based layer is adhesively bonded with BCB on top of the SOI circuit. The nanolasers are defined using e-beam lithography (EBL) on HSQ and patterned in the III-V layer with Inductively plasma etching (ICP). Chemical surface passivation based on Ammonium Sulfide [29] is employed in order to remove non radiative carrier recombination defects. A 1 µm thick layer of silica is deposited with Plasma-Enhanced Chemical Vapor Deposition (PECVD) in order to encapsulate the nanolasers. Finally, metallic nanowires to be used as heaters are made above the nanolasers using EBL, Ti/Au evaporation (10/60 nm), and lift-off.

**Derivation of the output directionality**

We develop the coupled mode analysis used to compute the waveguide output intensities and deduce the output contrast. We start from the rate equations describing two waveguide-coupled nanolasers:

$$j\frac{d\mathbf{A}}{dt} = \begin{pmatrix} \omega_A - j\Gamma_A & -j\mathrm{K} \\ -j\mathrm{K} & \omega_B - j\Gamma_B \end{pmatrix} \mathbf{A} + \begin{pmatrix} s_A \\ s_B \end{pmatrix} \quad (4)$$

where $s_A$ and $s_B$ are source terms which translate the spontaneous emission in the cavities. We deduce the stationary solutions by imposing $a_{A,B} = r_{A,B} e^{j\omega t}$ and solving Eq.4:

$$r_A = \frac{(\omega-\omega_B + j\Gamma_B)s_A + Ks_B}{(\omega-\omega_A + j\Gamma_A)(\omega-\omega_B + j\Gamma_B) - K^2} \quad (5)$$

$$r_B = \frac{(\omega-\omega_A + j\Gamma_A)s_B + Ks_A}{(\omega-\omega_A + j\Gamma_A)(\omega-\omega_B + j\Gamma_B) - K^2} \quad (6)$$

From which we compute the total output intensity and the output contrast, with in mind that $s_L = r_A + e^{j\phi}r_B$, and $s_R = e^{j\phi}r_A + r_B$. We compute $C_{\text{out}}$ assuming a linear growth of photon number below threshold, i.e. $s_{A,B} \propto \frac{P_{A,B}}{P_{A,B,t}}$. Here, $P_{A,B}$ is deduced from $\Gamma_{A,B}$, through the calibration functions. The mean loss-rate, $\overline{\Gamma}$, is taken constant and equal to $\Gamma_c/2$ in the colormap shown in Fig. 4A.

**Pump power-dependence of the total dissipation rate**

The amplitude decay rate $\Gamma_k$ of cavity $k$ can be split into three components:

$$\Gamma_k = \Gamma_{0,k} + \Gamma_c - \Gamma_{g,k} \quad (7)$$

where $\Gamma_{0,k}$ is the cavity intrinsic decay rate. $\Gamma_{g,k}$ is a gain term – this is why it is assigned to a minus sign – that depends on the carrier density, $N_k$, in the semiconductor and, by extension, to the optical power of the pump. We write

$$\Gamma_{g,k}(N_k) = \frac{V_a}{2} G(N_k) \text{ where } G(N_k) = G_0 \ln\left(\frac{N_k+N_s}{N_{tr}+N_s}\right) \qquad (8)$$

Here, $G_0$ is a gain factor that depends on the materials and cavity properties, $N_{tr}$ is the carrier density at transparency, and $N_s$ is a residual carrier density. The nanolaser wavelength is also affected by the carrier population which fills the electronic levels in the semiconductor, leading to a blue-shift of the cavity.

Below threshold, we assume a linear dependence of the carrier density with the pump power: $N_k = \varepsilon P$, prior to a clamping of the carrier density above threshold. Overall, the amplitude decay rate of cavity $k$ reads:

$$\Gamma_k(P) = \Gamma'_{0,k} + \Gamma_c - \frac{V_a}{2} G_0 \ln\left(\varepsilon \frac{P}{N_s} + 1\right) \text{ with } \Gamma'_{0,k} = \Gamma_{0,k} + \frac{V_a}{2} G_0 \ln\left(\frac{N_s}{N_s+N_{tr}}\right) \qquad (9)$$

## Acknowledgments

This work was partly supported by the French government through the France 2030 investment plan managed by the National Research Agency (ANR), as part of the Initiative of Excellence of Université Côte d'Azur under reference number ANR-15-IDEX-01, by the French RENATECH network and the ANR-22-PEEL-00010 BEP project.